\documentclass[fleqn,12pt]{article}
\usepackage{times,amsmath,amssymb,amscd}
\usepackage{theorem}
\theorembodyfont{\em} \theoremheaderfont{\bf}
\usepackage[dvips]{graphicx}
\usepackage{latexsym}
\makeatletter
\@addtoreset{equation}{section}

\makeatother

\begin{document}

%\tableofcontents
\title{Circular Stochastic Fluctuations in SIS Epidemics
with Heterogeneous Contacts Among Sub-populations}
\date{}

\author{{Jia-Zeng Wang$^{1}$\thanks{Corresponding author; Email: wangjiazen@yahoo.com.cn; Phone: (86)10-150-10720629.},
{Min Qian$^{2}$}, {Hong Qian$^3$}} \\
\small $^1$Department of Mathematics, Beijing Technology and Business University,\\
\small Beijing 100048, P.R.C.\\
\small $^2$School of Mathematical Sciences, Peking University, Beijing 100871, P.R.C.\\
\small $^3$Department of Applied Mathematics, University of
Washington,\\
\small Seattle, WA 98195, U.S.A. \\
} \maketitle

\centerline{\bf Abstract}

The conceptual difference between equilibrium and non-equilibrium
steady state (NESS) is well established in physics and chemistry.
This distinction, however, is not widely appreciated in dynamical descriptions of biological populations in terms of differential equations in which fixed point, steady state, and equilibrium are all synonymous.  We study NESS in a stochastic SIS
(susceptible-infectious-susceptible) system with heterogeneous individuals in their contact behavior represented in terms of
subgroups. In the infinite population limit, the stochastic dynamics
yields a system of deterministic evolution equations for population
densities; and for very large but finite system a diffusion process
is obtained. We report the emergence of a circular dynamics in the diffusion process, with an intrinsic frequency, near the endemic
steady state.  The endemic steady state is represented by a 
stable node in the deterministic dynamics; As a NESS phenomenon, 
the circular motion is caused by the intrinsic heterogeneity 
within the subgroups, leading to a broken symmetry and time irreversibility.

\vskip 0.3cm


%\noindent {\bf PACS number (s):} 89.75.-k, 05.70.Ln

\section{Introduction}

The mathematical description of population dynamics
is universally based on ``population change per unit time $=$ birth
rate $-$ death rate.'' \cite{murray_book,kot_book}. One of the most
widely employed nonlinear models for teaching population dynamics is
$\dot{n}=rn(1-n/C)$, where $r$ is the growth rate per capita when
the population is infinitely spars, and $C$ represents carrying
capacity. However, a careful reflection immediately leads to the
following question: Does $r(1-n/C)$ represent a decreasing per
capita birth rate; or is the birth rate being linear $rn$ and death
rate being $rn^2/K$? Such an explicit distinction has no consequence
in ordinary differential equation models. However, it has important
consequences to stochastically fluctuating dynamics of finite
populations \cite{qian_nonl_11}.

    The distinction is necessary when one develops
an individual behavior based stochastic model.  Population dynamics
in terms of stochastic, distributed individual behavior is a more
refined description of the biological reality.  The nonlinear
dynamics one often observes is a collective, emergent behavior at
the level of a population within which individuals making seemingly random choices with bias. In sociology and economics,
this type of modeling is called {\em agent-based}, and
in finance it is called {\em behavior finance}.

    A case in point is the infectious epidemics among a
large population with several behavioral subgroups.
It has long been recognized that various types of
heterogeneities play fundamentally different roles in
epidemics such as sexually transmitted diseases (STDs) \cite{Lajmanovich1976}.  Statistical and analytical studies
have defined and characterized heterogenous phenomena such
as the existence of core groups leading to the ``80/20 rule'' \cite{Woolhouse,Koopman}.

    The rise of ``network theory'' and its graph
representations have provided a new perspective on modeling
heterogeneity in population dynamics.  In particular, networks with
scale-free characteristics have been used to represent diverse
population interactions, such as human sexual contacts
\cite{Lil,Sch}, air transportations \cite{Collizza2006}, and the
World Wide Web \cite{Barabasi,Pastor3}. Dynamics of such networks
has attracted much attention \cite{AndersonRM,May,Pastor,Wang2007}.

    Another widely used approach to heterogeneity is to
introduce a probability distribution over a population. This
naturally leads to dynamic models based on stochastic processes. 
The stochastic approach has received growing attentions in recent years. For example, N\aa sell has extended the classical 
methods of moment-closure and pairwise approximation (see \cite{Nasell2003}  and the references cited
therein.)  Stochastic approaches have also been combined with
network dynamics
\cite{Collizza2006,Danger,Bauch,AndersonH,Hufnagel2004}.

In the present work, we consider the statistical
behavior, at the individuals level, which can itself be
``heterogeneous''.  In other words, each
subgroup has a different ``contact number''. This
notion is motivated, though no need to be, by
the {\em degree distribution} in the network theory.
Specifically, using nodes to represent individuals and
edges to represent contacts between individuals,
we envision every individual has a fixed number of
{\em half-edges} representing his/her level of social
activity.  The heterogeneity in our model, thus, comes
from the different number of half-edges an individual
possesses.  A contact between two individuals is
then represented by a connection of two half-edges.

Compared with relatively long time period of being infectious ($I$)
and susceptible ($S$), the act of disease transmission is often very
short, and will be considered as instantaneous.  Therefore, we shall
assume that the contact between two half-edges as a {\em collision}
event between two individuals.  Such contacts can lead to an
individual changing state from $S$ to $I$. In this way, our model is
different from the pairwise models which deal with concurrent
partnerships
\cite{Collizza2006,Danger,Bauch,AndersonH,Hufnagel2004}. There
exists at most one ``collision'' in the system at any time.

These considerations naturally lead to the  mass-action 
law, which we
adopt: With an identical rate, any two half-edges from different
individuals are possible to collide with equal probability. The
actual collision events are stochastic, happening in serial and
independently. This is known as a Poisson flow.  Furthermore, we
assume that the recovering time of an infectious individual obeys an
exponential distribution. Based on these assumptions, we define a
stochastic process on the level of population subgroups.
For a system with $K$ subgroups, the stochastic process is a
birth-death process with multi-dimensional discrete state
space $\mathbb{Z}^K$ \cite{Durrett94}.

In the limit of infinite population size, using the method of
$\Omega$-expansion \cite{Kampen2010}, we show that our stochastic
model becomes a system of deterministic evolution equations which
are closely related to the models previously studied by Lajmanovich
and Yorke \cite{Lajmanovich1976}, May and Lloyd \cite{May}, and others \cite{Pastor}.  This indicates that the assumptions
we made for the mechanism at the individual level 
are reasonable; and in
fact we have obtained a stochastic counterpart of the
classical model.  A stochastic model requires more explicit assumptions than those for differential equations.

As for the cause(s) of often observed noisy oscillations
(fluctuations) in epidemiological data, it is still controversial
\cite{Bjornstad}.  Deterministic framework focuses on interactions
between external forcing and inherent frequencies in nonlinear
dynamics \cite{Hethcote1991}.  Stochastic models, however,
illustrate a fundamental role of intrinsic randomness in the
patterns of disease cycles \cite{Dushoff,Alonso2006,Simoes2007}.
Such a debate echoes the nature of fluctuations/oscillations in the
concentrations of chemical species in a small volume. It has been
shown that in a stochastic, diffusion process model one can
unambiguously distinguish two types of mechanisms
\cite{qian_pnas_02}: stochastic fluctuations and nonlinear dynamic
complexity. Furthermore, it has been shown that nonlinear
oscillations in a stationary stochastic dynamics are intimately
related to a {\em probability circulations} associated with
time-irreversible Markov processes \cite{jqq,zqq,gqq}. One of the
objectives of the present work, thus, is to call for attentions to
population dynamic studies of chemical species which might serve as
a paradigm for dynamics of more complex systems
\cite{qian_nonl_11,wolynes}.

The interplay between stochasticity and nonlinearity in epidemics
was studied in \cite{Alonso2006}: an oscillatory spiral type steady
state in the deterministic system was shown to be amplified by the
demographic stochasticity. In our system, the oscillatory dynamics
is a consequence of finite population; it disappears in the
deterministic nonlinear dynamics.

Classical SIS systems with well-mixed homogenous individuals
have no oscillatory dynamics in either deterministic or stochastic
models. In the former, there exist at most two attractors:
a trivial stable node and a non-trivial stable node.
The term ``node'' here refers to a type of fixed point with
real eigenvalues in dynamical systems theory; it should
not to be confused with the same term in graph theory.
(A planar fixed point with complex eigenvalues is called a ``spiral''.)  Stochastic SIS model is a one-dimensional
birth-death process \cite{Reid}; it is known that such
a process is time-reversible if it is stationary.

Deterministic SIS dynamics with heterogenous
individuals is multi-dimensional; still it exhibits the same type
of behavior as the homogeneous
case \cite{Lajmanovich1976,Pastor}: All fixed points are nodes.
However, for the corresponding stochastic model in
this paper, we discover
that the multi-dimensional birth-death process can exhibit circular
dynamics. We shall first demonstrate this by a linear theory near the non-trivial stable node, i.e., via an Ornstein-Uhlenbeck (OU) process.
We then investigate the nonlinear, multi-dimensional
birth-death process and show the circular motion as a
fundamental property of the stochastic epidemic process.  It is
a consequence of the heterogeneity among individuals.

The paper is organized as follows. Sec. 2 describes the
stochastic contact process among individuals leading to the
definition of a multi-dimensional birth-death process.
In Sec. 3, we carry out the $\Omega$ expansion of van
Kampen \cite{Kampen2010} and obtain a diffusion approximation
of the multi-dimensional birth-death process in the limit
of large population size $N$.  The $\Omega$ expansion
consists of two steps which are generalizations of the Law of Large
Numbers and the Central Limit Theorem \cite{Ethier}.  In the first
step a system of ordinary differential equations (ODE) is obtained
under the $N^{-1}$ scaling.  Then conditioned on an
ODE solution and with a $N^{-1/2}$ scaling, an OU process is obtained \cite{qian_nonl_11,zhou_qian_11}.

In Sec. 4, the deterministic ODE system is analyzed.
We show under different conditions the system yields
either a positive non-trivial stable node or a
trivial one at the origin. The former corresponds to
an endemic state while the latter corresponds the
infection being eradicated.

Then in Sec. 5, the properties of the OU process near the
positive non-trivial stable node are studied.
While the stable node gives no indication of any oscillation by the ODE, the OU process undergoes a sustained
circular motion around the stable node | It is represented by a
nonzero stationary circular flux.

In Sec. 6, we use the simple system of two subgroups to
further illustrate the circular pattern of
infectious dynamics.  We show it is not a result of
$\Omega$-expansion approximation; we demonstrate
that the discrete multi-dimensional birth-death process
violates the so-called Kolmogorov cycle criterion.
Therefore, according to a mathematical theorem in irreversible
Markov processes, the circular flux exists in the discrete
model --- This feature originates from the heterogeneous
contacts and it does not occur in SIS systems with
homogenous populations.  The paper concludes with Sec. 7.

\section{The model}

\subsection{Individual contacts and recovery}

We consider total $N$ individuals in a population.
We assume that each
individual has a fixed ``contact degree'' in an epidemic, which is
represented by the number of {\em half-edges} (or valency)
associated to the individual.  We further assume that the entire
population can be partitioned into subgroups according to the
contact degrees of individuals. For example, groups 1, 2, 3, etc.,
represent individuals with contact degrees between $0-10$, $11-20$,
$21-30$, etc., respectively.  The rate of encounter between two
individuals from subgroups $i$ and $j$ is assumed to be proportional
to $i\times j$.

    Let $N_{k}$ be the number of individuals in the $k^{th}$
subgroup and total population $N=\sum_{k=1}^K N_k$.
We shall denote the fraction of the population
$N_{k}/N$ by $D_k$; $k=1,2,\cdots,K$, with $K$ being
finite.  In graph theoretical language, the $D_k$ is
the {\em degree distribution} for the contact network:
$\sum_{k=1}^K D_k=1$.

Following the standard compartmental modeling of epidemics, we
assume every individual in the population is in one of the two
states: susceptible (denoted by S) or infected (I). In the present
work, an individual becomes infectious immediately after being
infected.  Let $T_i$ denotes the infectious period of individual
$i$, and we assume that all $T_i$; $i=1,2,\cdots,N$, are independent
and identically distributed following an exponential dwell time with
mean value $1$.  An infectious individual becomes `susceptible' once
again when his/her infectious period is terminated. These
assumptions imply the heterogeneity we consider is in the contacts
between individuals, while the infected individuals behave
statistically homogeneous.

    One can visualize the contacts among individuals in our
model as follows:  There are regular, repeated ``touching'' or
``collisions'' within pairs of half-edges, which trigger one of the
individuals with certain probability to change its state. Every
``touching'' is instantaneous.  If one of the individual is
infectious and the other is susceptible, then an infection can
occur.  In some sense, the dynamics is no different from an
autocatalytic chemical reaction in aqueous solution $A+X\rightarrow
2X$, as in the celebrated Lotka-Volterra model \cite{lotka_1910}.
See \cite{qian_nonl_11} for a discussion on dynamic
isomorphisms between cellular biochemical and epidemiological population dynamics.

\subsection{Defining a stochastic epidemics}

We now turn the previous verbal description into a mathematical
model.  We denote the stochastic demographic process
\begin{equation}
    X(t) =
    \left(I^{(1)}(t),\cdots,I^{(k)}(t),
    \cdots,I^{(K)}(t)\right), \ \ t\ge 0,
\end{equation}
in which $I^{(k)}(t)$ is the number of infected individuals in the
$k^{th}$ subgroup at time $t$, and accordingly $N_{k}-I^{(k)}(t)$
will be the number of susceptible individuals.  $X(t)$
is a continuous-time, $K$-dimensional birth-death process.
Since $0\leq I^{(k)}(t)\leq N_{k}$, $X(t)$ takes values in a
bounded subset of the $K$-dimensional lattice $\mathbb{Z}^{K}$.

When initially there are infectious individuals in a
population, there will be a stochastic epidemic process.
Let $\lambda$ be the rate of infection for a single
half-edge. $\lambda/(\sum_{m=1}^{K}mN_{m})$ then is
normalized by the total number of half-edges in the system
such that in the infinite population limit,
$N_m, \lambda \rightarrow\infty$ but the ratio is unchanged.
Then, a give half-edge makes pair with an infectious one
at the rate of
\begin{equation}
    \widetilde{\lambda} =
    \left(\frac{\lambda}{\sum_{m=1}^{K}mN_{m}}\right)
        \sum_{m=1}^{K}mI^{(m)}.
\end{equation}
For a given susceptible individual with degree $k$, it will
be `infected' at the rate of $k\widetilde{\lambda}$.
Thus the number of susceptible individuals $N_k-I^{(k)}(t)$
decreases by $1$ in the $k^{th}$ subgroup with rate $k(N_{k}-I^{(k)})\widetilde{\lambda}$.
Illustratively, the transitions and corresponding rates of the
process $X(t)$ are given as follows:
\begin{equation}
\begin{array}{cc}
\mbox{\underline{transitions:}} \pm e_{k} & \mbox{\underline{rates}}  \\
    +e_{k}=(\underbrace{0,\cdots,0,+1}_{k},0,\cdots,0) &  \ \
    J_{+}^{k}= {\lambda}k(N_k-I^{(k)})
            \left(\frac{\sum_{m=1}^{K}mI^{(m)}}
            {\sum_{m=1}^{K}mN_{m}}\right) \\
    -e_{k}=(\underbrace{0,\cdots,0,-1}_{k},0,\cdots,0) &  J_{-}^{k}= I^{(k)} \\
\end{array}
\label{tranrat}
\end{equation}
Here $k=1,2,\cdots,K$. Recall that without loss of
generality the recovery rate of an infected individual is
1.

Denotes the probability $P(\vec{\rho};t) =
\Pr\{X(t)=\vec{\rho}\}$, $\vec{\rho}\in\mathbb{Z}^K$.
The stochastic dynamical
system is characterized by the master equation for
the probability distribution
\begin{subequations}
\begin{eqnarray}
\frac{d}{dt}P(\vec{0},t) & = &
\sum_{k=1}^{K}J_{-}^{k}(\vec{0}+e_{k})P(\vec{0}+e_{k}),
\\
\frac{d}{dt}P(\vec{\rho},t)&=&
    \sum_{k=1}^{K}J_{+}^{k}(\vec{\rho}-e_{k})P(\vec{\rho}-     e_{k})+\sum_{k=1}^{K}J_{-}^{k}(\vec{\rho}
        +e_{k})P(\vec{\rho}+e_{k})
\nonumber\\
  &&
  -\sum_{k=1}^{K}\left(J_{-}^{k}(\vec{\rho})+J_{+}^{k}
         (\vec{\rho})\right)P(\vec{\rho}), \ \ \
  \  \left(\vec{0}<\vec{\rho}<\vec{N_k}\right)
\\
\frac{d}{dt}P(\vec{N_k},t) & = &
\sum_{k=1}^{K}J_{+}^{k}(\vec{N_k}-e_{k})P(\vec{N_k}-e_{k})-\sum_{k=1}^{K}J_{-}^{k}(\vec{N_k})P(\vec{N_k}).
\nonumber\\[-12pt]
\end{eqnarray}
\label{masterE}
\end{subequations}
It should be noted that the solution to Eq. (\ref{masterE})
with initial data
$P(\vec{\rho};0)=\delta(\vec{\rho}-\vec{\rho_1})$ is the
transition probability $\Pr\{\vec{\rho}(t)=\vec{\rho}|\vec{\rho}(0)=\vec{\rho}_1\}$ for the Markov process $X(t)$.

It is obvious that the $\vec{0}$ is an absorbing state. This means
the infinite long-time behavior of the system is always the {\em
extinction} of the infectious population \cite{Nasell2001}.  We are,
however, interested in the pre-extinction dynamics.  In many
biological population dynamics, extinction is too long a time scale
to be relevant. To explore the pre-extinction dynamics, we shall
first consider large population ($N_k$) limit and establish the corresponding nonlinear deterministic behavior in the following two sections.
Pre-extinction dynamics can also be studied in terms
of a transient landscape \cite{zhou_qian_11}.  However,
we shall not pursue this approach in the present paper.

\section{Population-size ($\Omega$) expansion for large $N$}

    With increase population size in an increasing
space, the population dynamics for discrete individuals can
be described by a continuous variable representing the
``density'', or ``frequency'', or any other intensive quantity.  The
Law of Large Numbers from the theory probability then
dictates the dynamics of the system approaches to a
deterministic behavior.  Mathematically, a systematic method
called $\Omega$ expansion is widely used in
statistical physics \cite{Kampen2010}.  See its recent
application in epidemiological modeling \cite{Alonso2006}.

Let us first introduce the notation of a ``step operator" $\mathbb{E}_{k}$ \cite{Kampen2010}, which represents the
elementary changes of the discrete $X(t)$.  It is
defined by its transformation of a function $f(\vec{\rho})$:
\begin{equation}
\mathbb{E}_{k}f(\vec{\rho})\triangleq f(\vec{\rho}+e_{k})
 \ \textrm{ and } \
\mathbb{E}_{k}^{-1}f(\vec{\rho})\triangleq f(\vec{\rho}-e_{k}).
\end{equation}
Then the master equation in Eq. (\ref{masterE}) can be
rewritten in a much more compact form
\begin{eqnarray}
\label{gen_master}
\frac{d}{dt}P(\vec{\rho};t)=
\sum_{k=1}^{K}\left(\mathbb{E}_{k}-1\right)J_{-}^{k}(\vec{\rho})P(\vec{\rho})+\sum_{k=1}^{K}\left(\mathbb{E}_{k}^{-1}-1\right)J_{+}^{k}(\vec{\rho})P(\vec{\rho}).
\end{eqnarray}
Now we introduce a new vector $\vec{y}$, which scales as $N^{-1}$ so
represents the density vector of infected population, and a
fluctuation $\vec{\xi}$ | which scales as $N^{-1/2}$ such that
\begin{equation}
            \rho_{k}=Ny_{k}+N^{1/2}\xi_{k}
                              + o\left(N^{1/2}\right).
\label{vk_expansion}
\end{equation}
Accordingly the distribution is now written as a function of
$\vec{\xi}$,
\begin{equation}
           P(\vec{\rho};t)=\Pi(\vec{\xi},t),
\end{equation}
and the master equation (\ref{masterE}) in the new variable
takes the form
\begin{eqnarray}
&& \frac{\partial\Pi}{\partial t} - \sum_{k}
N^{1/2}\frac{dy_{k}}{dt}\frac{\partial \Pi}{\partial \xi_{k}} \nonumber\\
&=& \sum_{k=1}^{K}\left(-N^{-1/2}\frac{\partial}{\partial
\xi_{k}}+\frac{1}{2}N^{-1}\frac{\partial^{2}}{\partial\xi_{k}^{2}}+\cdots\right)\left(\lambda
kN(D_k-y_k)+\lambda kN^{1/2}\xi_{k}\right)\nonumber\\
&& \left(\frac{\sum_{l}l(y_k+N^{-1/2}\xi_{l})}{\langle k\rangle}\right)\Pi(\vec{\xi})\nonumber\\
&+& \sum_{k=1}^{K}\left(N^{-1/2}\frac{\partial}{\partial
\xi_{k}}+\frac{1}{2}N^{-1}\frac{\partial^{2}}{\partial\xi_{k}^{2}}+\cdots\right)\left(Ny_{k}+N^{1/2}\xi_{k}\right)\Pi(\vec{\xi}),
\label{omegaexp}
\end{eqnarray}
in which parameter
\begin{equation}
               \langle k\rangle = \frac{\sum_{k=1}^K k N_k}
                  {\sum_{k=1}^K N_k} =\sum_{k=1}^K k D_k.
\end{equation}
The conditions for the terms of order $N^{1/2}$ to vanish are
\begin{equation}
\frac{dy_{k}(t)}{dt} = -y_{k}(t)+\frac{\lambda k}{\langle
k\rangle}\left(D_k-y_{k}(t)\right)\sum_{j=1}^{K}jy_{j}(t),
   \ \ k=1,2,\cdots,K.
\label{deq}
\end{equation}
This is a system of $K$-coupled nonlinear ordinary differential
equations (ODEs) for $y_k(t)$: The deterministic nonlinear
dynamics for infinite size population.  The system of ODEs
is a generalization of the one-dimensional logistic equation $\dot{y}=-y+\lambda(D-y)y$,
which has two fixed points, one at 0 and another positive one at $D-1/\lambda$ when $\lambda D > 1$.
This result should be understood as the Law of Large Numbers
for the stochastic process $X(t)$.

The terms of order $N^{0}$ in (\ref{omegaexp}) yield
\begin{eqnarray}
\frac{\partial \Pi(\xi_k,t)}{\partial t} &=&
-\sum_{k,l}\left(\frac{\lambda kl(D_k-y_k)}{\langle
k\rangle}-\delta_{kl}\left(\lambda k\frac{\sum jy_{j}}{\langle
k\rangle}+1\right)\right)\frac{\partial}{\partial\xi_{k}}\left(\xi_{l}\Pi(\vec{\xi},t)\right)
\nonumber \\
&+&\frac{1}{2}\sum_{k}\left(\lambda k(D_k-y_k)\frac{\sum
jy_{j}}{\langle
k\rangle}+y_k\right)\frac{\partial^{2}\Pi(\vec{\xi},t)}
  {\partial\xi_{k}^2}.
\label{FPK}
\end{eqnarray}
This is a
time-inhomogeneous linearly diffusion process centered at the
time-dependent $\vec{y}$.  It should be understood as the
counterpart of the central limit theorem for $X(t)$.

Eqs. (\ref{vk_expansion}), (\ref{deq}) and (\ref{FPK}) together show
that the evolution of the stochastic dynamical system described by
master equation (\ref{gen_master}), when the population size is
large but not infinite, can be characterized by two parts: One
represents a deterministic nonlinear dynamics of the densities, $y_k(t)$, in
infinite population-size limit, and another represents the
fluctuations in terms of stochastic diffusion, with density function
$\Pi(\xi_k,t)$, centered  around the deterministic trajectory.  It
is important to point out that the $y_k(t)$ is the behavior of an
infinitely larger population; it is {\em not} in general the mean
dynamics for a finite population with nonlinear interaction.  Mean
dynamics of a finite population is of course ``deterministic'', but
usually it can not be described by a self-contained set of
autonomous nonlinear differential equations. See \cite{mom_cl}
for a recent study on the moment closure problem.

\section{Deterministic dynamics with a stable node}

We now study the system of ODEs (\ref{deq}).  It should be noted
that similar equations have been proposed phenomenologically in the
past \cite{Lajmanovich1976,AndersonRM,Pastor}. In particular, the
elegant mathematical analysis in \cite{Lajmanovich1976} has dealt
with a more general class of problems, which can easily be applied
here. In order to make the material more accessible to readers with
less mathematical backgrounds, here we present a simplified
recapitulation of the earlier work for the particular ODE system
(\ref{deq}). Our method is also more explicit; it can be conveniently
adopted in numerical computations for the non-trivial stable
steady state: Using the $C$ and $G(C)$ defined in Eqs. (\ref{eq_4.1}) and (\ref{the_GC}), one can easily obtain
$C^*$ from arbitrary chosen initial $C$,
with computational iterations with rapid convergence.
For more rigorous mathematical treatment, however, one
is referred to \cite{Lajmanovich1976}.

Let
\begin{equation}
  C(t)\triangleq \frac{\sum_{m=1}^{K}my_{m}(t)}{\langle k\rangle},
\label{eq_4.1}
\end{equation}
where $0\leq C(t) \leq 1 $ since $\sum_{m=1}^{K}mD_{m}=\langle
k\rangle$ and $D_{m}\geq y_{m}(t)$. $y_{m}$ is the density of
infectious individuals with degree $m$, so $C(t)$ represents the
mean fraction of infectious half-edges in the system.

According to Eq. (\ref{deq}), the positive equilibrium is located
at $y^*_k=\frac{\lambda k D_k C^*}{1+\lambda k C^*}; k=1,2,\cdots, K$. If $y_k(t)<\frac{\lambda k D_k C}{1+\lambda k C}$, $y_k(t)$
will increase, so does $C(t)$.

Even though the dynamics is $K$-dimensional, it is easier to work
with the single variable $C(t)$.  We introduce a function of $C$:
\begin{equation}
G(C)\triangleq  \frac{1}{\langle
k\rangle}\sum_{m=1}^{K}m\frac{\lambda m D_{m}C}{1+\lambda m C}.
\label{the_GC}
\end{equation}
It can be explained as the averaged infectious half-edges caused by
$C$ in an unit time. We see that $C(t)$ increases if
\begin{equation}
C(t)=\frac{\sum_{m=1}^{K}my_{m}(t)}{\langle k\rangle}<G(C).
\end{equation}
Alternatively, $C(t)$ decreases when $C>G(C)$.  The auxiliary
function $G(C)$, therefore, has a fixed point $C^*\neq 0$ as the
solution of $G(C)=C$.  Accordingly the steady states of
Eq. (\ref{deq}), $y^*_k=\frac{\lambda k D_k C^*}{1+\lambda k
C^*}$, can be obtained.

\begin{figure}[h]
%\begin{minipage}[t]{10cm}
\includegraphics[width=12cm] {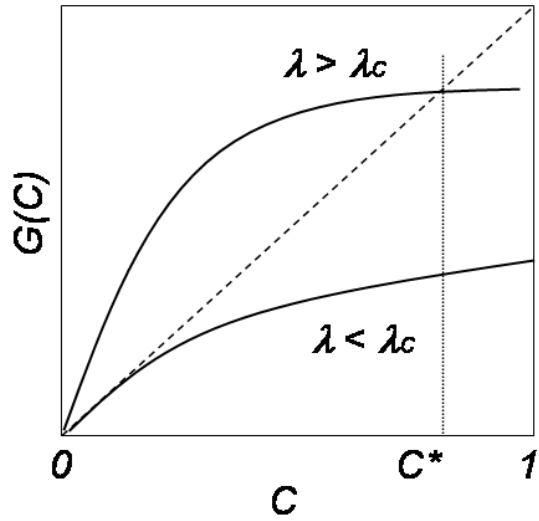}
\caption {\small The functions $G(C)$, given in Eq. (\ref{the_GC}),
with two different values of $\lambda$. When $\lambda>\lambda_c$,
there will be a positive, stable node corresponding to an endemic
steady state with a sustained infection.} \label{FunGc}
%\end{minipage}
\end{figure}

The fixed point $C^*$ is determined by the properties of the
function $G(C)$.  We note that $G(0)=0$, and
\begin{equation}
G(1)=\frac{1}{\langle k\rangle}\sum_{m}^{K}\frac{\lambda
m^2D_{m}}{1+\lambda m }=\frac{1}{\langle k\rangle}\sum_m^K m D_{m}
\frac{\lambda m}{1+\lambda m}\leq \frac{1}{\langle k\rangle}\sum_m^K
m D_{m}=1.
\end{equation}
Furthermore, we have $G(C)$'s derivative:
\begin{equation}
G'(C)=\frac{1}{\langle k\rangle}\sum_{m}^{K}\frac{\lambda m^2
D_{m}}{(1+\lambda m C)^2}>0,
\end{equation}
and $G''(C)<0$. All together, we see that function
$G(C)$ is monotonically increasing and concave in the
region $0\leq C \leq 1$.  Fig. \ref{FunGc} shows the
function $G(C)$.

There are two cases according to the values of
$\lambda$.  When $\lambda<\lambda_c=\langle k\rangle/\langle
k^2\rangle$, $G'(0)<1$.  Then there is only one fixed point
$C^{*}=0$ and correspondingly $y^*_k=0$ $\forall k$.  It is
stable.  In this case, the disease will fade out
eventually.

If $\lambda>\lambda_c$, $G'(0)>1$, the fixed point $\vec{0}$ becomes
unstable. Simultaneously there emerges a unique stable fixed point
$0<C^{*}<1$, and accordingly $y^{*}_{k}=\frac{\lambda k
C^*D_{k}}{1+\lambda k C^{*}}$.  In this case, there will be an
endemic state.  The results can be seen directly from Fig.
\ref{FunGc}.  There is a trans-critical bifurcation at
$\lambda=\lambda_c$, as in the logistic equation
$\dot{y}=-y+\lambda(D-y)y$ when $\lambda=1/D$.

\subsection{Linear analysis of steady states}

One can linearize equation (\ref{deq}) near its steady
states and obtain
\begin{equation}
\frac{d\vec{y}}{dt}=B\vec{y},
\label{lineardeq}
\end{equation}
in which the linear community matrix $B$ has elements
\begin{equation}
         B_{kl}(\vec{y^*})=\frac{\lambda
kl(D_k-y_k^{*})}{\langle k\rangle}-\delta_{kl}\left(\lambda
k\frac{\sum_{j=1}^K jy_{j}^{*}}{\langle k\rangle}+1\right). \label{Bkl}
\end{equation}

At the non-trivial, internal steady state, one can
further obtain the complete eigenvalues of $B$:
\begin{equation}
         \Lambda_1=-\lambda C^*,  \ \
         \Lambda_k=-\lambda k C^*-1; \ \ 2\leq k\leq K.
\end{equation}
Therefore this steady state is a locally stable {\em node}.
This means that there is no any indication of oscillatory
behavior, at this deterministic level, near the steady
state.

At the steady state for extinction: $\vec{y^*}=\vec{0}$, we have
$B_{kl}=\lambda klD_k/\langle k\rangle-\delta_{kl}$ from
(\ref{Bkl}), which yields the eigenvalues
\begin{equation}
           \Lambda_1=\lambda\frac{\sum k^2 D_k}
             {\langle k\rangle}-1, \ \
          \Lambda_i=-1; \ \ 2 \leq i\leq K.
\end{equation}
The eigenvector corresponding to the first eigenvalue
$\Lambda_1$ is
\begin{equation}
\left\{1, \underbrace{\cdots,\ \frac{iD_i}{D_1\langle k^2\rangle}
        \left(\sum_{k\neq i}^K k^2D_k+D_i\right)>0, \cdots}_{i=2,3,\cdots, K}
    \right\}.
\end{equation}
In the neighborhood of $\vec{0}$, this eigenvector locates within
the first quadrant. It is stable when $\lambda\langle
k^2\rangle/\langle k\rangle<1$, and unstable when $\lambda\langle
k^2\rangle/\langle k\rangle>1$.

\section{Non-equilibrium fluctuations near the positive stable node}

    We now turn our attention to stochastic dynamics
near the stable internal steady state $\vec{y}=\vec{y^*}$. Note that
we have shown through the above linear analysis that
$\vec{y^*}$ is a node, with all
eigenvalues being real.  To consider stochastic fluctuations, we
consider Eq. (\ref{FPK}). Substituting the deterministic steady
state solution into Eq. (\ref{FPK}), it is reduced to a
time-homogeneous Fokker-Planck equation defined on the entire
$\mathbb{R}^K$
\begin{equation}
\frac{\partial \Pi(\vec{\xi},t)}{\partial t}=\nabla\cdot\left(\frac{A}{2}\nabla
\Pi(\vec{\xi},t)-B\vec{\xi}\ \Pi(\vec{\xi},t)\right), \label{linearFPE}
\end{equation}
whose stationary solution is called Ornstein-Uhlenbeck (OU) process.

For the trivial stable node, the corresponding Fokker-Planck
equation is only defined on the first quadrant.  This leads to a
singular diffusion equation, which is outside the scope of
the present paper and will be the subject of a separated study.

    The constant drift matrix in Eq. (\ref{linearFPE}) is
precisely the community matrix $B$ in Eq. (\ref{lineardeq}). The
diffusion tensor is a diagonal matrix $A_{kk}(\vec{y^*})=2y_{k}^*$
since we have the useful relationship $\frac{\lambda k}{\langle
k\rangle}(D_k-y_k)\sum jy_{j}^*=y_{k}^*$ from Eq. (\ref{deq}). The
stochastic trajectories of OU process defined by Eq.
(\ref{linearFPE}) follows a linear stochastic differential equation:
\begin{equation}
\frac{d\vec{\xi}}{dt}=B\vec{\xi}(t)+\Gamma \vec{\zeta}(t),
\end{equation}
where $\Gamma\Gamma^T=A$, $\vec{\zeta}(t)$ is a Gaussian white noise
with $E[\vec{\zeta}(t)\vec{\zeta}^{T}(t')]=\delta(t-t')I$, and $I$
is the identity matrix.

If one sets the initial condition
\begin{equation}
        \Pi(\xi,0) = \prod_{i=1}^K \delta\left(\xi_i-\xi_i^o\right),
\end{equation}
Then the fundamental solution, i.e., Green's function, to Eq.
(\ref{linearFPE}) is a multivariate Gaussian distribution
\cite{foxron,qian_prsa_01}.
\begin{equation}
          \Pi(\vec{\xi},t) = \Pi_0\exp \left\{-\frac{1}{2}
    \left(\vec{\xi}-\vec{\mu}(t) \right)^T\Xi(t)^{-1}\left(\vec{\xi}-\vec{\mu}(t)\right)
               \right\},
\label{gs}
\end{equation}
where $\Pi_0$ is a normalization factor.
The first and second order moments satisfy
\cite{Kampen2010}:
\begin{equation}
\frac{d\mu_k(t)}{dt} =\sum_{j}B_{kj}\mu_k,
  \ \ \
\frac{d\Xi(t)}{dt}=B\Xi+\Xi B^T+A. \label{eqFSD}
\end{equation}
The solutions to Eq. (\ref{eqFSD}) are:
\begin{equation}
   \vec{\mu}(t)=e^{tB}\vec{\xi}_{0},   \ \ \
\Xi(t)=\int_{0}^{t}e^{(t-t')B}Ae^{(t-t')B^{T}}dt'.
\end{equation}
Since the eigenvalues of $B$ are all negative, we have ${\langle
\vec{\xi}\rangle}(\infty)=\vec{0}$ and $\Xi(t)$ approaches to a
$\Xi^s$ as the solution to Eq. (\ref{basic1}):
\begin{equation}
    B\Xi^s+\Xi^sB^T = -A.
\label{basic1}
\end{equation}
Equation (\ref{basic1}) is analogue to the well-known Einstein's
fluctuation-dissipation relation, in which $A$ is the covariance of
the fluctuating white noise, $B$ is the dissipative linear
relaxation rates, and $\Xi^s$ is the equilibrium covariance.
In thermal physics, $\Xi^s$ is proportional to $k_BT$.

Very interestingly, we note that $A^{-1}B$ is not symmetric. This
implies the stationary process has certain breaking symmetry with
respect to time reversal \cite{qian_prsa_01}. In physics, there is a
refined distinction between an equilibrium and a non-equilibrium
steady state. Such a distinction is not widely appreciated in
dynamical descriptions of biological populations in terms of
differential equations in which fixed point, steady state, and
equilibrium are all synonymous.

According to the method previously developed in physics
\cite{qian_nonl_11,qian_prsa_01,KAT}, there exists a stationary
circular current density
\begin{equation}
    \vec{j}^{(c)} = (\Phi\vec{\xi})\Pi^s(\vec{\xi}),
\label{cflow}
\end{equation}
in which $\Phi$ is a $K\times K$ matrix with zero
trace and $\vec{j}^{(c)}$ is a divergence free
vector field $\nabla\cdot\vec{j}^{(c)}=0$.
The linear force $B\vec{\xi}$ near the steady state,
then, can be decomposed into two orthogonal parts
\begin{equation}
   B= -\frac{1}{2}A\Xi^{-1}+\Phi.
\label{basic2}
\end{equation}
In Eq. (\ref{basic2}), the first term
$-\frac{1}{2}A\Xi^{-1}\vec{\xi}$ is a conservative
force with a potential function $U(\vec{\xi})=$ $\frac{1}{2}\vec{\xi}^T\left(\Xi^s\right)^{-1}\vec{\xi}$.
It generates a motion towards the origin. In another words,
it represents a {\em stability landscape}
$\textsl{U}(\vec{\xi})$.  The second term
$\Phi\vec{\xi}$ generates a stationary circular motion on the
level set of $\textsl{U}$.  To see this,  we observe the
orthogonality
\begin{equation}
       \vec{j}^{(c)} \perp \nabla \Pi(\vec{\xi})
\end{equation}
due to $\vec{\xi}^T\Xi^{-1}\Phi\vec{\xi}=0$. The stationary
probability distribution does not change along the direction of
$\vec{j}^{(c)}$. The maintenance of the stationary distribution
comes from two parts: one is the detailed along the gradient of
$U(\vec{\xi})$, and another is circular along its level curve.

\begin{figure}[h]
%\begin{minipage}[t]{20cm}
\includegraphics[width=6.75cm]{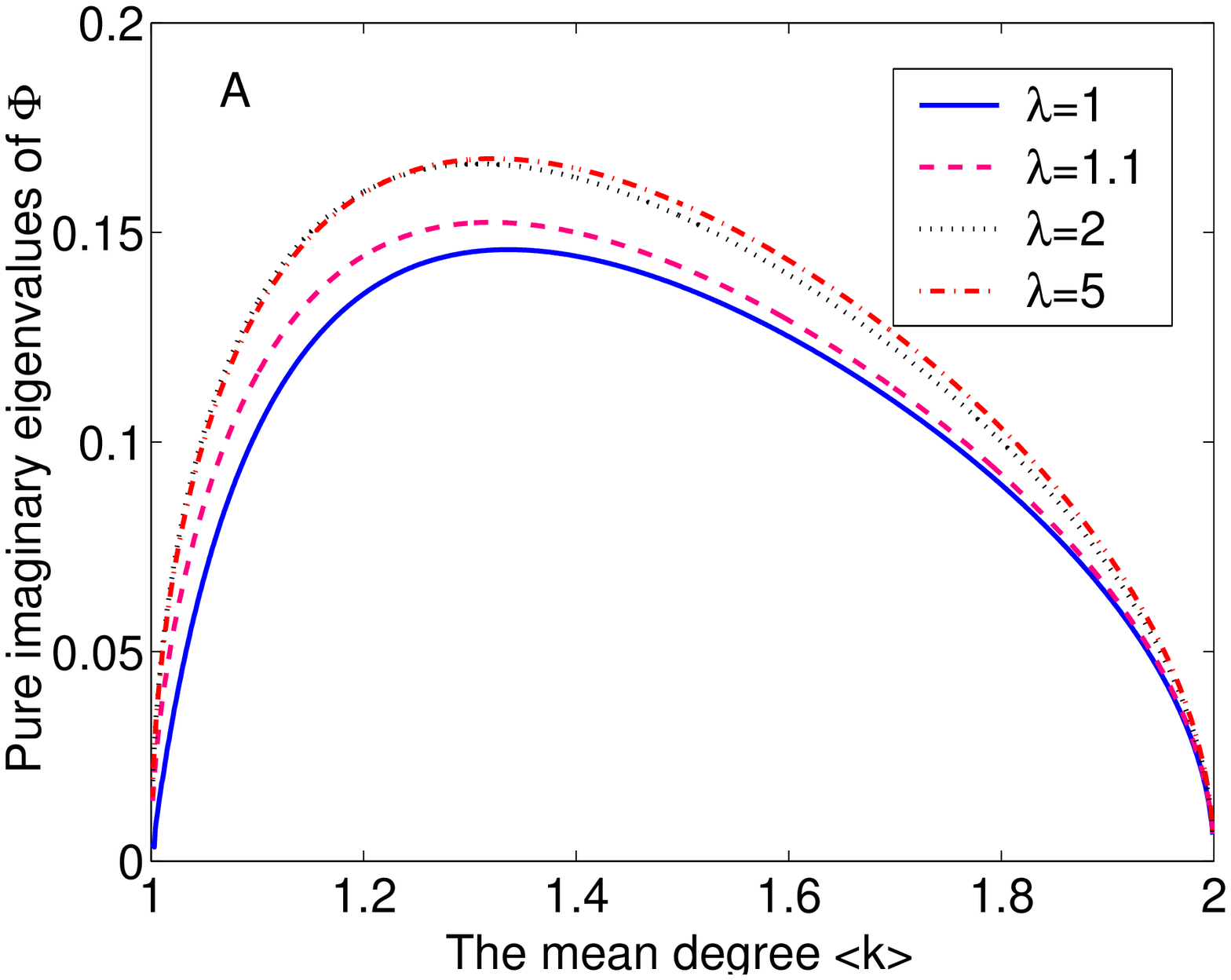}
\includegraphics[width=6.75cm]{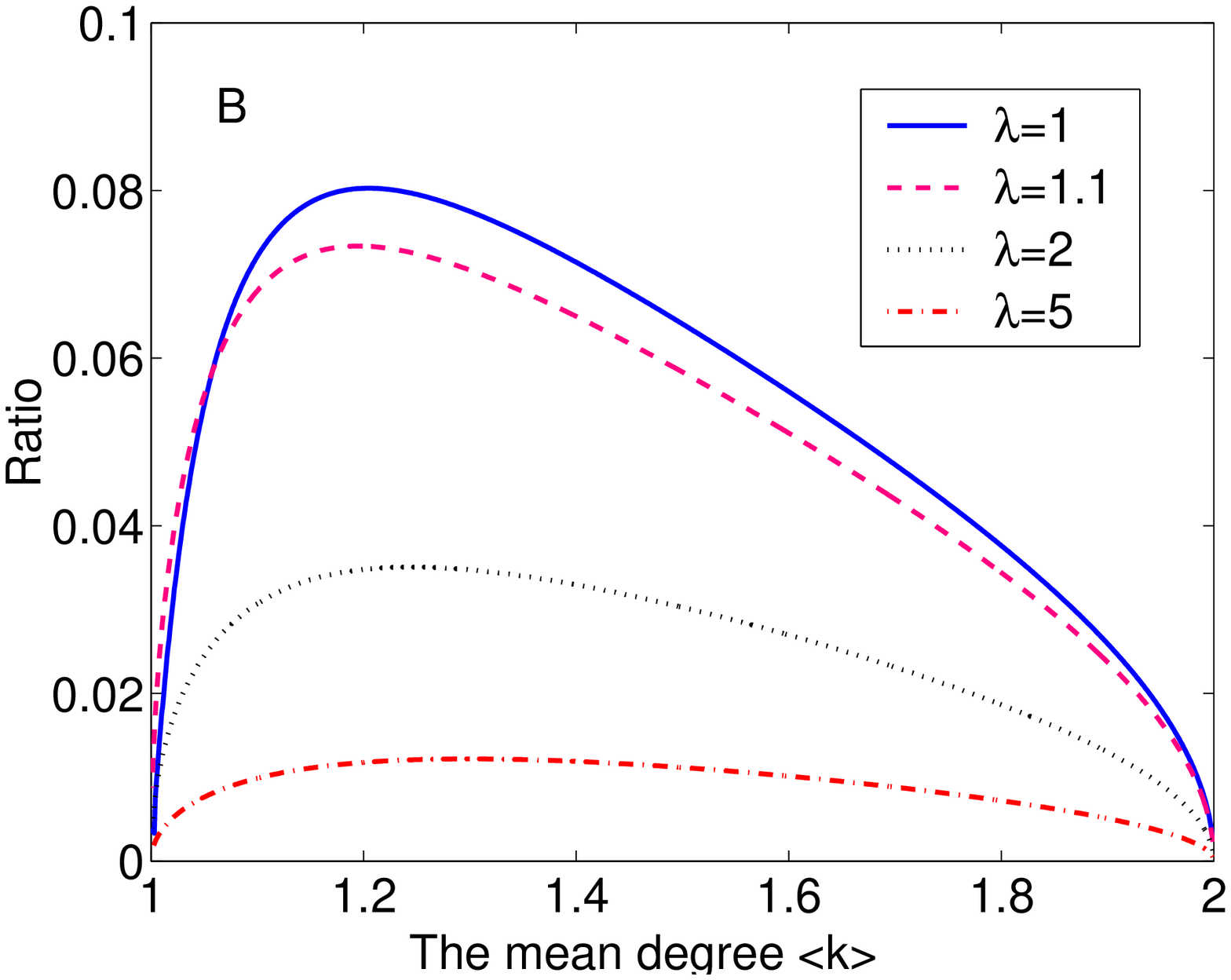}
\caption {\small (A) For $K=2$, the intrinsic
frequence, i.e., the pure imaginary eigenvalues of
$\Phi$, as a function of $\langle k\rangle$ and $\lambda$.
$\langle k\rangle$ changes from
1.999 to 1.001 due to $D_1$ changes from $1/1000$ to
$999/1000$ , and the corresponding $D_2=1-D_1$ changes
from $999/1000$ to $1/1000$.
(B): the IR (imaginary-to-real) ratio, the ratio
of the imaginary eigenvalues of $\Phi$ to the
sums of two real eigenvalues of matrix $B$, as a function
of $\langle k\rangle$ and $\lambda$.  Note the
critical $\lambda_c=1$ here; hence when $\lambda<1$
the endemic steady state disappears.}
\label{eigen_phi}
%\end{minipage}
\end{figure}

    For heterogeneous SIS with two subgroups, i.e., $K=2$ and $k=1,\ 2$,
the matrix $B$ from Eq. (\ref{Bkl}) is
\begin{equation}
  B = \left(\begin{array}{cc}
         \frac{\lambda(D_1-2y_1^*-2y_2^*)}{D_1+2D_2}-1 &
         \frac{2\lambda(D_1-y_1^*)}{D_1+2D_2} \\[5pt]
         \frac{2\lambda(D_2-y_2^*)}{D_1+2D_2} &
         \frac{2\lambda(2D_2-y_1^*-4y_2^*)}{D_1+2D_2} -1
    \end{array}\right),
\end{equation}
and matrix
\begin{equation}
   A = \left(\begin{array}{cc}
            2y_1^* & 0 \\ 0  & 2y_2^*
     \end{array}\right),
\end{equation}
in which $y_2^*\in (0,D_2]$ is a root of
\begin{equation}
  4\left(y_2^*\right)^2-\left(2D_1+12D_2-
    \frac{\langle k\rangle}{\lambda}\right)y_2^* +
      2D_2\left(D_1+4D_2-\frac{\langle k\rangle}{\lambda}
         \right) = 0,
\end{equation}
and
\begin{equation}
    y_1^* = \frac{\langle k\rangle y_2^*}{2\lambda
            (D_2-y_2^*)} - 2y_2^*.
\end{equation}
Therefore, with parameters $\lambda=1$ and
$D_1=D_2=\frac{1}{2}$, thus $\langle k\rangle = 1.5$,
the matrix
\begin{equation}
    \Phi = \left(\begin{array}{cc}
                -0.0644  &  0.1204    \\   -0.1903  &  0.0644
               \end{array}\right).
\end{equation}
The eigenvalues of this $\Phi$ is pair of pure imaginary
numbers $\pm 0.137i$.   This reflects an {\em intrinsic
frequency} of the dynamics.  Fig. \ref{eigen_phi}A shows how the
frequency changes with $\langle k\rangle$ for several different
values of $\lambda$. For heterogeneous populations with two
subgroups $N_1$ and $N_2$, the maximal effect of heterogeneity has
to be $\langle k\rangle$ between 1 and 2; where the imaginary
eigenvalues of $\Phi$ is also the largest.

    One of the widely used technique to capture and
characterize oscillatory dynamics is the method of power spectrum
\cite{zqq}.  For a strong oscillatory motion with noise, its power
spectrum exhibits an off-zero peak at the intrinsic frequency.  By
``strong'', we mean the intrinsic frequency has to be significantly
greater than the relaxation rate of the stochastic dynamics.
\cite{qq_prl_00} has shown that the ratio of imaginary part to real
part of an eigenvalue has to be at least greater than
$\frac{1}{\sqrt{3}}=0.577$.  Fig. \ref{eigen_phi}B shows the ratio
between the intrinsic frequency, given by the imaginary eigenvalue
of $\Phi$, and the relaxation rate, given by the eigenvalues of the
community matrix $B$.  This explains why one does not observe an
off-zero peak in the power spectra of the simulated dynamics (data
not shown).

\section{Intrinsic circular dynamics in multi-dimensional
birth-death processes}

    We now address a crucial question unanswered in Sec. 5:
Whether the novel circular dynamics near the endemic
steady state, shown by the OU approximation, is a consequence
of the population-size expansion approximation, or is it
a general result for the finite population birth-death
process.  The answer is affirmative.  To illustrate this,
we shall again consider the special case of $K=2$ and
analyze the planar system in some detail.  The same methods and
results can easily be generalized to any $K>2$, but the algebra will be more cumbersome.

    First, we need to address the issue of
long-time behavior of the original birth-death process. In Sec. 4 we
have shown that when $\lambda>\lambda_c$, the deterministic dynamics
has a stable internal node representing an endemic steady state,
while the extinction state $y_i=0$ $(1\le i\le K)$ is unstable.  On
the other hand, it had been shown in Sec. 2 that the infinite
long-time behavior of the original birth-death process is always
{\em extinction}, even though it might take extremely long time.
This disparity between the stable deterministic steady state and the
stochastic long-time behavior indicates a separation of two
different time scales: The time scale on which the infectious
dynamics reaches a stationary pattern among different populations,
and the time scale for extinction of the infectious population
\cite{Nasell2001,vellela_qian,zhou_wu_ge}.  In the light of the
time-scale separation, this is a well-understood subject in
population dynamics.  There are in fact two fundamentally
different types of {\em extinction}: one that occurs as a
consequence of nonlinear dynamic, and another that occurs as
rare events that is impossible according to nonlinear
dynamics \cite{zhou_qian_11}.

    To study the ``long-time'' dynamics in the
pre-extinction phase, one of the methods of attack is {\em
quasi-stationary approximation}
\cite{Nasell2001,zhou_qian_11,vellela_qian}. Conceptually, we only
consider the stationary probability distribution conditioned on the
{\em survival probability} at time $t$.  To carry out the
computation, one can introduce a very tiny transition probability
$\varepsilon\ll 1$ representing the process can ``return from'' the
state $\vec{0}$.  This approximation abolishes the absorbing state.
Since the total number of states is finite for our original
birth-death process and it is irreducible, an unique stationary
distribution exits. Biologically, the $\varepsilon$ can be
interpreted as an infinitesimal invasion by migration, or
recurrence, of infectious individuals. The time evolution of
probability in (\ref{masterE}a) then becomes
\begin{equation}
\frac{d}{dt}P(\vec{0},t) =
\sum_{k=1}^{K}J_{-}^{k}(\vec{0}+e_{k})P(\vec{0}+e_{k})-\varepsilon\delta(P(\vec{0},t)\in[1-\varepsilon,1]).
\end{equation}

\begin{figure}[h]
%\begin{minipage}[t]{20cm}
\includegraphics[width=12cm]{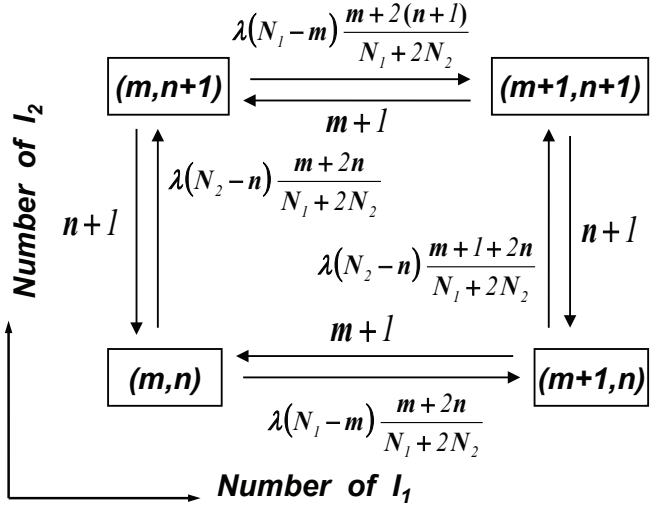}
\caption {\small The ``phase plane'' for a two-dimensional
birth-death process $\left(I_1,I_2\right)$ where $I_1$ and $I_2$
take non-negative integers.  All the possible transitions among four
neighboring states are shown as the arrows and the corresponding
rates are labeled by the transitions ({\em i.e.}, Eq.
(\ref{tranrat})).} \label{MEGfigA}
%\end{minipage}
\end{figure}

    In the case of $K=2$, the probability transition
rates in the planar system are illustrated in Fig. \ref{MEGfigA}.
Such a diagram is known as a {\em master equation graph} in
stochastic chemical kinetics \cite{Beard2008,bishop_qian}.  It is
useful tool in ``visualizing'' the dynamics of the master equation
in Eq. (\ref{masterE}), on a par with the phase portrait of planar
nonlinear dynamics.

\begin{figure}[h]
%\begin{minipage}[t]{20cm}
\includegraphics[width=12cm] {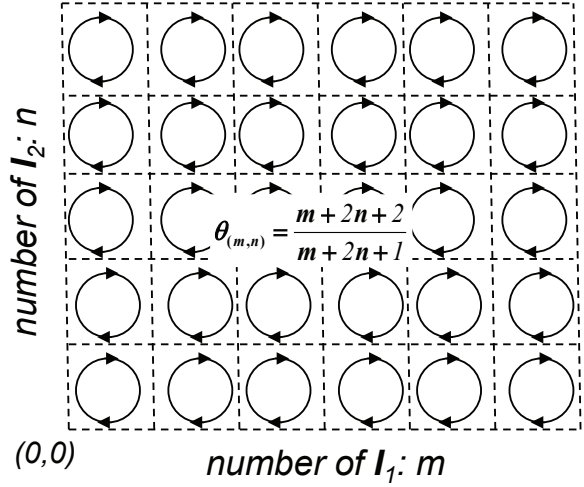}
\caption {\small Stationary birth-death process can have a
hidden circular dynamics, analogous to the vortex in a fluids
\cite{qian_prl}. The existence of stationary circulation can be
determined by the Kolmogorov cycle number $\theta_{(m,n)}$.  For our
model with $K=2$, the $\theta_{(m,n)}=\frac{m+2n+2}{m+2n+1}$ and the
circulation is always clockwise.  For a larger, arbitrary closed
circular path, the value of $\theta$ is the product of all the
individual $\theta$'s within. A stationary Markov process is
time-reversible if and only if all the $\theta$'s are 1.  In
mathematical statistics and in statistical physics, this is also
known as detailed balance.} \label{MEGfigB}
%\end{minipage}
\end{figure}

According to a key theorem in the theory of irreversible Markov
chain, a sufficient and necessary condition for the existence
of circular flux in the stationary process is the Kolmogorov
cycle condition \cite{jqq,zqq}.
In Fig. \ref{MEGfigA} the values for all the
transition rates around a square cycle are given. Let $J_{+}^{(m,n)}$
denote the ``clockwise" circular rate product
\begin{equation}
   J_{+}^{(m,n)} = q_{(m,n)\rightarrow(m,n+1)}q_{(m,n+1)
    \rightarrow(m+1,n+1)}q_{(m+1,n+1)\rightarrow(m+1,n)}
    q_{(m+1,n)\rightarrow(m,n)},
\end{equation}
in which the $q_{(m,n)\rightarrow(m,n+1)}$ is the transition rate
from grid point $(m,n)$ to grid point $(m,n+1)$.  Similarly
$J_{-}^{(m,n)}$ is the ``counterclockwise" rate product
\begin{equation}
    J_{-}^{(m,n)} = q_{(m,n+1)\rightarrow(m,n)}q_{(m+1,n+1)
    \rightarrow(m,n+1)}q_{(m+1,n)\rightarrow(m+1,n+1)}
    q_{(m,n)\rightarrow(m+1,n)}.
\end{equation}
Then the Kolmogorov cycle criterion $\theta$ for each
little square cycle is:
\begin{equation}
    \theta_{(m,n)}=\frac{J_{+}^{(m,n)}}{J_{-}^{(m,n)}}.
\end{equation}
Substituting the values for the transition rates in Fig.
\ref{MEGfigA}, we have $\theta_{(m,n)}=\frac{m+2n+2}{m+2n+1}$, shown
in Fig. \ref{MEGfigB}.
We note that $\theta_{(m,n)}$ is very close to $1$ for
large $m$ or $n$. So generally the circular motion is weak.
This explain why there is no circular motion in the
deterministic dynamics of densities, when
$N\rightarrow \infty$.

    The $\theta_{(m,n)}$s in Fig. \ref{MEGfigB} can be
heuristically understood as ``vortex'' in a vector field. They are
the fundamental units of circular dynamics.  For a larger closed
circular path, its $\theta$ is simply the product of all the
$\theta$'s within. The significance of the $\theta$ value lies in
the theory of stationary, irreversible Markov processes.  According
to the cycle decomposition theorem for irreversible Markov processes
\cite{jqq}, the $\theta_{(m,n)}$ is the ratio of numbers of
occurrence, following the stationary, stochastic birth-death
 process, of the clockwise cycle
$(m,n)\rightarrow(m,n+1)\rightarrow(m+1,n+1)\rightarrow(m+1,n)\rightarrow(m,n)$
to that of the corresponding counter-clockwise cycle. A stationary
process is time-reversible if and only if each and every cycle has
$\theta=1$.

The Fig. \ref{MEGfigB} gives the insight that the infection dynamics
is ``moving clockwise" in the phase plane in its stationary state.
We know that a one-dimensional stationary birth-death process is
always time-reversible. Therefore, irreversibility in the present
example comes from the {\em heterogeneity} in the infectious
population.  This is a novel dynamical behavior emerged in
population with heterogeneous structures.

\begin{figure}[h]
\includegraphics[width=12cm] {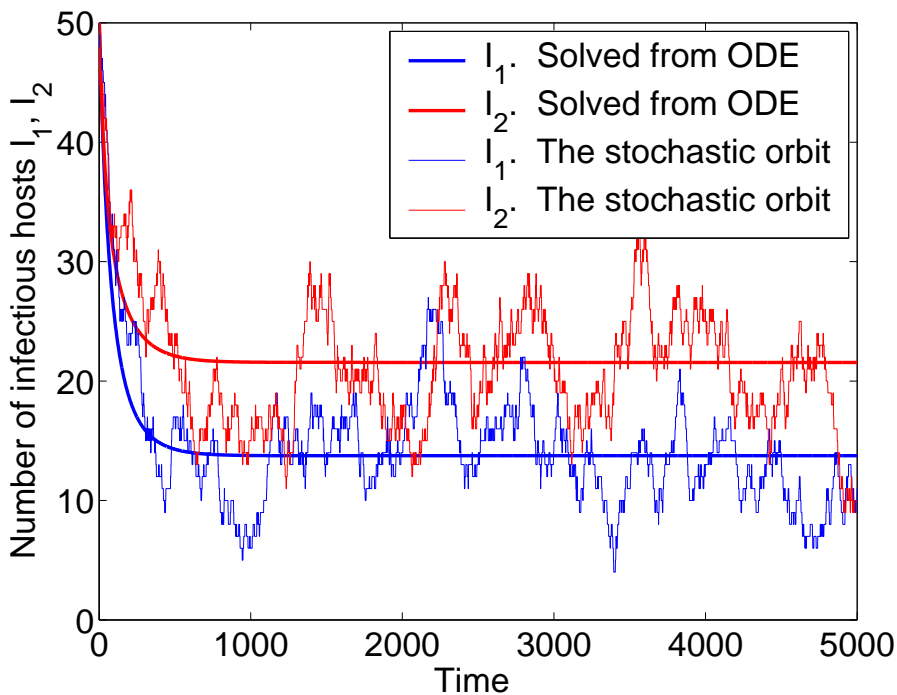}
\caption {\small A typical stochastic realization of the SIS model.
Here we considered two subgroups with $k=1,2$. The parameters used
in simulation are $N_1=50$, $N_2=50$ (thus $\langle k\rangle =
1.5$), and the transmission rate $\lambda=1$. The simulation is
carried out according to Gillespie's algorithm \cite{Gillespie}. The
thickened lines are orbits of the deterministic equations
(\ref{deq}) with same $\lambda$ and $D_1=D_2=\frac{1}{2}$.}
\label{orbit}
\end{figure}

In order to see things more clearly, we carried out stochastic
simulations for the stochastic process.  Firstly we consider two
subgroups with $k=1,\ 2$.  Other parameters used are $N_1=50$,
$N_2=50$ and the transmission rate $\lambda=1$.  The initial
conditions are chosen randomly.  Fig. \ref{orbit} shows the typical
trajectories of the two infecting population sizes, $I_1$ and $I_2$.
For comparison, the two thick lines are the solutions to the
corresponding deterministic equations (\ref{deq}): $y_1(t)\times N$
and $y_2(t)\times N$, respectively.

\section{Conclusion and discussion}

    This paper has two intertwined threads.  First, it
reports a novel observation that for an
SIS epidemic dynamics in a finite population with
heterogeneous subgroups, there is
an emergent inherent frequency,
even though the corresponding deterministic
nonlinear dynamics for infinite population shows no
indication whatsoever.  This result provides new
insight to the understanding of fluctuations in
epidemiological data.

    Since the phenomenon is at the juncture
between dynamics in terms of systems of deterministic
ordinary differential equations (ODEs) and
in terms of stochastic birth-death processes,
the relation between these two types of
mathematical models is rigorously investigated.
We want to emphasize that in our approach, the
nonlinear dynamics is an emergent, collective
phenomenon of the stochastic population dynamics.
The second thread of the paper, therefore, is to advance
a systematic approach to nonlinear population
dynamics based on individual's behavior with
uncertainties, which can be represented in terms of
probability distributions.  From this starting point, the
traditionally employed ODEs can be more explicitly
justified as the limiting behavior of infinitely large
population for a nonlinear stochastic dynamics, called
Delbr\"{u}ck-Gillespie process in cellular biochemistry
\cite{qian_nonl_11,bishop_qian}.  Furthermore,
a Gaussian like stochastic
process can also be derived to account for the
population fluctuations in the large, but
finite, population.

Indeed, on the level of individuals, epidemic processes
are essentially stochastic, and the heterogeneity are
inevitable.  In the present study, we distinguish
the stochastic, but statistically identical behavior
(i.e., within each subgroup) from statistically non-identical
behavior (i.e., between different subgroups).
We show a more careful model building based ``mechanisms''
of the infection, though still quite crude, nevertheless
can provide further insights into the dynamics on the
population level.

One might wonder why the circular dynamics disappears in
deterministic system.  The reason lies in the equality
$\theta_{(m,n)}=\frac{m+2n+2}{m+2n+1}$. As we have shown, the ODE
dynamics equations are obtained from stochastic processes via the
Law of Large Numbers, which is in order of $N^{-1}$, where $N$ is
the total number of individuals in the population.  In the limit of
$N\rightarrow\infty$, $\theta_{(m,n)}\rightarrow 1$.

In summary, we proposed a microscopic, statistically heterogeneous
contact process for individual-based epidemiological modeling.  The
fundamental stochastic events are the instantaneous ``contacts"
between two individuals and the recovery of an infectious individual.  Within an infinitesimal time interval, these events
are sequential and independent.  The stochastic process is a
Poisson flow.
We applied this approach to two-state SIS dynamics. Following the
method of $\Omega$ expansion, we obtained a system of deterministic
ODEs for the densities for a large population, as well as a linear
diffusion system near the stable steady state.

There are two time scales in the dynamics of SIS epidemics.
Compared with the time in which the infection reaches a ``stationary
pattern'' among the subgroups, the time for ultimate extinction
is very long when $\lambda>\lambda_c$ \cite{zhou_qian_11}.
We used the method of quasi-stationary approximation to
study the long-time behavior on the ``short time scale''.

Another assumption adopted here is the mass action law
which requires well mixing. In population dynamic term,
it is assumed that individuals are moving rapidly within a
relatively small, highly ``fluid'' community.  In a large
spatial scales, the contacts and spatial movements
involve geographic factors; then a spatial model is required.
In the applications of epidemic dynamic models to 
``virus infection'' on the World Wide Web, however, this 
restriction is hardly there.

\section{Acknowledgements}

JZW acknowledges support from China Postdoctoral Science 
Foundation (Special Grade No. 201003021), 
Beijing Municipal Government Fund for Talents (2011,
No. D005003000009), and the NSFC (No. 11001004). 
HQ thanks T. C. Reluga for helpful comments.

{}
\end{document}